# A procedure to correct for target thickness effects in heavy-ion PIXE at MeV energies

Alessandro Zucchiatti[1*], Patricia Galán[2] and José Emilio Prieto[2,3]

[1]*Universidad Autónoma de Madrid, 28049 Madrid, Spain*

[2]*Centro de Micro Análisis de Materiales (CMAM), Universidad Autónoma de Madrid, 28049 Madrid, Spain*

[3]*Dpto. de Física de la Materia Condensada, IFIMAC and Instituto ¨Nicolás Cabrera¨, Universidad Autónoma de Madrid, 28049 Madrid, Spain*

ABSTRACT

We describe a novel procedure for the calculation of correction factors for taking into account the effect of target thickness to be applied to the determination of cross sections of X-ray emission induced by heavy ions at MeV energies. We discuss the origin of the correction and describe the calculations, based on simple polynomial fits of both the theoretical cross sections and the ion energy losses. The procedure can be easily implemented. We show several examples for a set of targets specifically produced for cross section measurements and for various combinations of ion type and energy.

*Corresponding author: Alessandro Zucchiatti. Universidad Autónoma de Madrid, Rectorado, calle de Einstein 3, 28049 Madrid, Spain. Tel. +34914975529, e-mail alessandro.zucchiatti@uam.es

## 1. INTRODUCTION

The high potential of heavy-ions induced X-ray emission (HI-PIXE), combined with Secondary Ion Mass Spectrometry (SIMS) at MeV energies for materials characterization has been recently highlighted in several implemented facilities [1-5]. This topic has been the subject of a (still active) coordinated research project (CRP) promoted by the International Atomic Energy Agency (IAEA) [6]. The preliminary CRP work consisted in a literature review [7] that pointed-out the scarcity of data for both ionization and X-ray production cross sections. The details of this review are expected to be published in the CRP final report. There is thus an interest in increasing the required data basis in terms of both the ion species considered and the materials analyzed. New data on X-ray production cross sections will be, on the one hand, a convenient, reliable operative tool and, on the other hand, an experimental basis for the testing of theoretical models.

Measuring cross-sections induced by heavy ions is more complex than for the case of protons or alpha particles, which are routinely used for materials characterization. The use of heavy ions presents several problems [8]. A relevant one is the need for consideration of energy losses caused by effects of finite target thicknesses. The specific energy loss of a particle in a material depends essentially on the inverse of the square of its velocity. Therefore, an ion of a given energy crossing a target will experience a higher energy loss when compared with a proton or an alpha particle of the same energy. The difference increases with the ion mass. When extracting [9] cross sections from fundamental parameters, like the detector solid angle and efficiency, the number of incident ions, the reaction yield and the target thickness, one needs to be sure that the ion energy has been reduced only by a negligible amount compared to the incoming energy, otherwise the result must be corrected for the energy change of the ion across the target. Only if this is done, the measured yield can be safely assigned to a cross-section at a well determined energy. The problem becomes more relevant by the difficulty of fabricating targets which are thin enough to produce negligible energy losses, even for a limited range of low ion masses.

A solution to this problem consists of a suitable calculation of a global factor to account for target thickness effects. In this paper we describe the development and testing of a rather simple calculation method to correct cross section values for X-ray production obtained from fundamental parameters within a thin target formalism.

## 2. TARGET THICKNESS EFFECTS IN THE MEASUREMENT OF HI-PIXE X-RAY PRODUCTION CROSS SECTIONS

Let us assume that a target of mass number A, bombarded by a given ion species, is thin enough in order not to induce sizeable changes in the involved physical quantities. The emission of X-rays can then be described by the following equation:

$$Y_X(E_I, \vartheta) = N_I \cdot \frac{N_0 \rho dx}{A} \cdot \varepsilon_{abs}(E_X) \cdot 4\pi \cdot \frac{d\sigma}{d\Omega}(E_I, \theta) \cdot e^{-\mu(E_X)\rho dx} \qquad (1)$$

Here $Y_X(E_I, \theta)$ is the X-ray yield observed at an angle $\theta$ with respect to the beam direction, $E_I$ is the incoming ion energy, $N_I$ is the number of incident ions, $\rho \cdot dx$ is the target areal density, $N_0$ is the Avogadro number and $d\sigma/d\Omega(E_I, \theta)$ is the reaction cross section. $\varepsilon_{abs}(E_X)$ and $\mu(E_X)$ are the detector absolute efficiency and the target attenuation coefficient for X-rays of energy $E_X$, respectively.

If the quantities defining the experimental set-up and the cross section are known and the yield is measured, the equation above can be used for extracting the target areal density. This is its typical analytical application aimed at determining the quantity of an element in a sample. Alternatively, it can be used to extract the production cross section from the measured yield if the target thickness (areal density) and composition are known. In both cases, equation (1) is valid if: *a)* there is practically no difference between the production cross sections at the ion entrance ($E_I$) and exit ($E_f$) energies; *b)* the attenuation of the X-rays in their path from the production point to the target surface, accounted for by the exponential term in equation (1), is also negligible. In all the practical cases considered in this work the exponential term can be assumed equal to be equal to 1 (see Table I). If the above conditions are not fulfilled, it is necessary to integrate equation (1) between the entrance and exit ion energies, obtaining:

$$Y_X(E_I, \vartheta) = N_I \cdot \frac{N_0}{A} \cdot \varepsilon_{abs}(E_X) \cdot 4\pi \cdot \int_{E_f}^{E_I} \frac{\frac{d\sigma}{d\Omega}(E, \vartheta)}{S(E)} dE \qquad (2)$$

If we assume that in the energy interval $E_I$-$E_f$, the cross section is constant and equal to $\left(\frac{d\sigma}{d\Omega}\right)_{meas}$, we may re-write equation (2) as:

$$Y_X(E_I, \vartheta) = N_I \cdot \frac{N_0}{A} \cdot \varepsilon_{abs}(E_X) \cdot 4\pi \cdot \left(\frac{d\sigma}{d\Omega}\right)_{meas} \int_{E_f}^{E_I} \frac{1}{S(E)} dE \qquad (3)$$

From equations (2) and (3) one deduces that the cross section measured from the yield is:

$$\left(\frac{d\sigma}{d\Omega}\right)_{meas} = \int_{E_f}^{E_I} \frac{\frac{d\sigma}{d\Omega}(E, \vartheta)}{S(E)} dE \bigg/ \int_{E_f}^{E_I} \frac{1}{S(E)} dE \qquad (4)$$

The measured cross sections will be linked to the cross section at the incoming ion energy by a factor $F$ given by:

$$F = \frac{\left(\frac{d\sigma}{d\Omega}\right)_{meas}}{\frac{d\sigma}{d\Omega}(E_I, \theta)} = \frac{1}{\frac{d\sigma}{d\Omega}(E_I, \theta)} \int_{E_f}^{E_I} \frac{\frac{d\sigma}{d\Omega}(E, \vartheta)}{S(E)} dE \bigg/ \int_{E_f}^{E_I} \frac{1}{S(E)} dE \qquad (5)$$

The *F* factor depends on the variation of both the cross section and the ion energy loss with energy, to an extent that depends on the ion type, the target material and the target thickness. In the case of HI-PIXE, we expect the cross section to show no resonances and to increase monotonically with energy **[10-11]**. This makes the calculation of the integrals in equation (5) quite simple and implies that the *F* factor will be always less than 1. The cross section, determined using equation (3), will have to be increased accordingly.

### 3. CALCULATION OF CORRECTION FACTORS

Analytical formulas for the correction factor have been given in literature **[12,13]** and have been used to check that corrections are minimal **[14]**, often of the same order of magnitude or lower than experimental errors. As an alternative procedure, we propose to calculate *F* by fitting with simple functions the cross sections and the energy losses and solving numerically both integrals in equation (5). We have verified that this is a fast procedure by applying it to the measurement of X-ray production on the targets and for the ion-energy combinations listed in Table I. Target thicknesses have been determined from the analysis of elastically backscattered 2.0 MeV alpha particles **[9]** performed with the help of the SIMNRA program **[15]**. The values obtained and their errors are reported in Table I. By taking into consideration several examples of theory-experiment comparison, as published in literature **[9, 14, 16-21]**, we conclude that the theoretical ECPSSR **[10,11]** curves represent reasonably well the dependence on energy of the experimental cross sections. Assuming then that the theoretical and experimental cross section only differ by a constant factor *K*, this will cancel out in equation (5). Using the code SRIM **[22]**, we have computed the energy loss of ions and their ranges in the used targets as a function of energy. Also range curves have been computed for all the incoming beam energies listed in table I and used to extract, by linear interpolation, the corresponding beam exit energies $E_f$. It is important to notice that, at the lowest beam energies, the energy loss amounts at most to 3.8% percent of the incoming ion energy for all the targets considered except for $Fe_3O_4$ (the thickest one), for which the loss is of the order of 10-12 %. It is possible to fit both the energy loss function *S(E)* and the differential cross section $d\sigma/d\Omega(E)$ with simple and easy-to-handle polynomial functions of $3^{rd}$ order at most, as shown by the examples of figure 1 and 2. In all cases, we were able to cover practically all the energies listed in Table I by fitting *S(E)* in at most three and $d\sigma/d\Omega(E)$ in at most four different energy ranges. Doing so, we have achieved inside each region a better fit than that obtained in the same region by a single fit over the entire energy range. For carbon, the worst fit of S(E) shows a value $R^2$ = 0.99639 corresponding to the case of the $Ta_2O_5$ target in the range 5-9 MeV, while all ECPSSR cross section fits show a value $R^2$ = 1. Carbon values are reported in Table II together with the equivalent results for the other ions. The calculation of the two definite integrals in equation (5) can be done using the Simpson rule:

$$\int_{E_f}^{E_I} f(E)dE = \frac{E_I - E_F}{6}\left[f(E_I) + f(E_F) + 4f\left(\frac{E_I + E_F}{2}\right)\right] \qquad (6)$$

where *f(E)* is a polynomial function of either four (for *S(E)*) or eight (for *dσ/dΩ(E)/S(E)*) parameters, some of which may be equal to zero. The approximation is very good as proved by the matching (Table I) of the experimentally determined target thicknesses and the values obtained from the integral of equation (3) considering all ions. The calculations are implemented in a worksheet and are quite straightforward. As selected examples, figure 3 shows the *F* factor calculated for the thickest target ($Fe_3O_4$) and all the ion-energy combinations used, while the results of a calculation for a heavier incoming ion (Br) and all targets is given in figure 4. The ratio of the measured cross section to the true one increases with energy, as expected, and decreases with the ion *Z* number. With the exception of the $Fe_3O_4$ target, the minimum values, at the lowest beam energies, amount to 92% for C ions of 4 MeV, 89% for Si ions of 8 MeV, 89% for Cl ions of 9 MeV, 85% for Cu ions of 10 MeV and 80% for Br ions of 10 MeV. In many cases the correction factor turns out to be of the same order of the experimental uncertainty budget (errors on target thickness, charge accumulation, detector efficiency, fitting procedures…) or lower.

4. **CONCLUSIONS**

We have shown, for a complete set of ion-target energy combinations, that the calculation of the correction factor for the effect of finite target thickness in the determination of heavy-ion induced X-ray emission cross sections can be easily and accurately performed. The energy loss curves can be accurately fitted using polynomials of at most 3$^{rd}$ order within convenient energy regions to cover the required intervals. The correction factor depends on the behavior of the cross section with energy. For the case of heavy-ion PIXE, the ECPSSR theory predicts a monotonic increase of the X-ray production cross section with energy. This behavior can be fitted as well with polynomial functions of at most 3$^{rd}$ order in wide enough energy ranges. The procedure can be implemented in a worksheet in a simple, efficient and reliable way.

**Acknowledgements**

This work was performed within the IAEA coordinated research project (CRP) #F11019: "Development of molecular concentration mapping techniques using MeV focused ion beams" and has been partly supported by Project No. MAT2014-52477-C5-5-P of the Spanish MINECO.

**TABLES**

| Targets | Analyte Thickness [µg/cm$^2$] | | | Target Thickness [µg/cm$^2$] | Density [g/cm$^3$] | Maximum Analyte X-ray attenuation | Target Thickness from eq. (3) [µg/cm$^2$] | Ions | Energies [MeV] |
|---|---|---|---|---|---|---|---|---|---|
| TiN/Si | 11.2 | ± | 0.4 | 14.52 | 4.82 | 0.998 | 14.5 ± 0.5 | C | 4,6,9,12,15,20 |
| TiN-C/Si | 19.1 | ± | 0.5 | 27.97 | 4.82 | 0.999 | 28 ± 1 | Si | 8,12,16,20,24,28 |
| TiO$_2$/Si | 11.56 | ± | 0.2 | 19.38 | 4.23 | 0.997 | 19.4 ± 0.5 | Cl | 9,12,19,26,33,40 |
| ZnO/Si | 22.9 | ± | 0.4 | 29 | 5.61 | 0.995 | 29.2 ± 0.8 | Cu | 10,15,20,25,30,35,40 |
| Fe$_3$O$_4$-C/Si | 64 | ± | 2 | 85.76 | 5 | 0.999 | 87 ± 2 | Br | 10,15,20,25,30,35,40 |
| RuO$_2$/Sigradure | 11 | ± | 0.2 | 15.57 | 6.97 | 1.000 | 15.7 ± 0.4 | | |
| Ta$_2$O$_5$/Sigradure | 15.9 | ± | 0.4 | 19.78 | 8.18 | 1.000 | 20.2 ± 0.5 | | |
| Nb$_2$O$_5$/sigradure | 14.3 | ± | 0.2 | 21.05 | 4.6 | 0.997 | 21.2 ± 0.5 | | |

**Table I**
Left: List of the targets, showing the analyte thickness, both as determined by using 2 MeV alpha particle backscattering and the cumulative result from eq. (3). Right: List of the ions and energies considered in the calculation of target effects corrections applying the procedure described in the text.

| Ion | S(E) fitting | | | ECPSSR fitting | | |
|---|---|---|---|---|---|---|
| | worst R2 | Target | Energy Range [MeV] | worst R2 | Target | Energy Range [MeV] |
| C | 0.99639 | Ta2O5 | 5-9 | | | |
| Si | 0.99273 | RuO2 | 20-29 | 0.99978 | TiO2 | 10-13 |
| Cl | 0.9883 | ZnO | 30-40 | 0.99997 | Ta2O5 | 38-40 |
| Cu | 0.99994 | RuO2 | 7-20 | 0.99989 | ZnO | 8-10 |
| Br | 0.99997 | Nb2O5 | 20-30 | 0.99997 | Nb2O5 | 13-20 |

**Table II** Illustration of the quality of the fittings performed. The worst $R^2$ obtained for each ion-target-energy combination is listed. Results are shown for the fitting of the energy dependence of the specific energy loss (left) and for the ECPSSR cross sections (right).

**FIGURES**

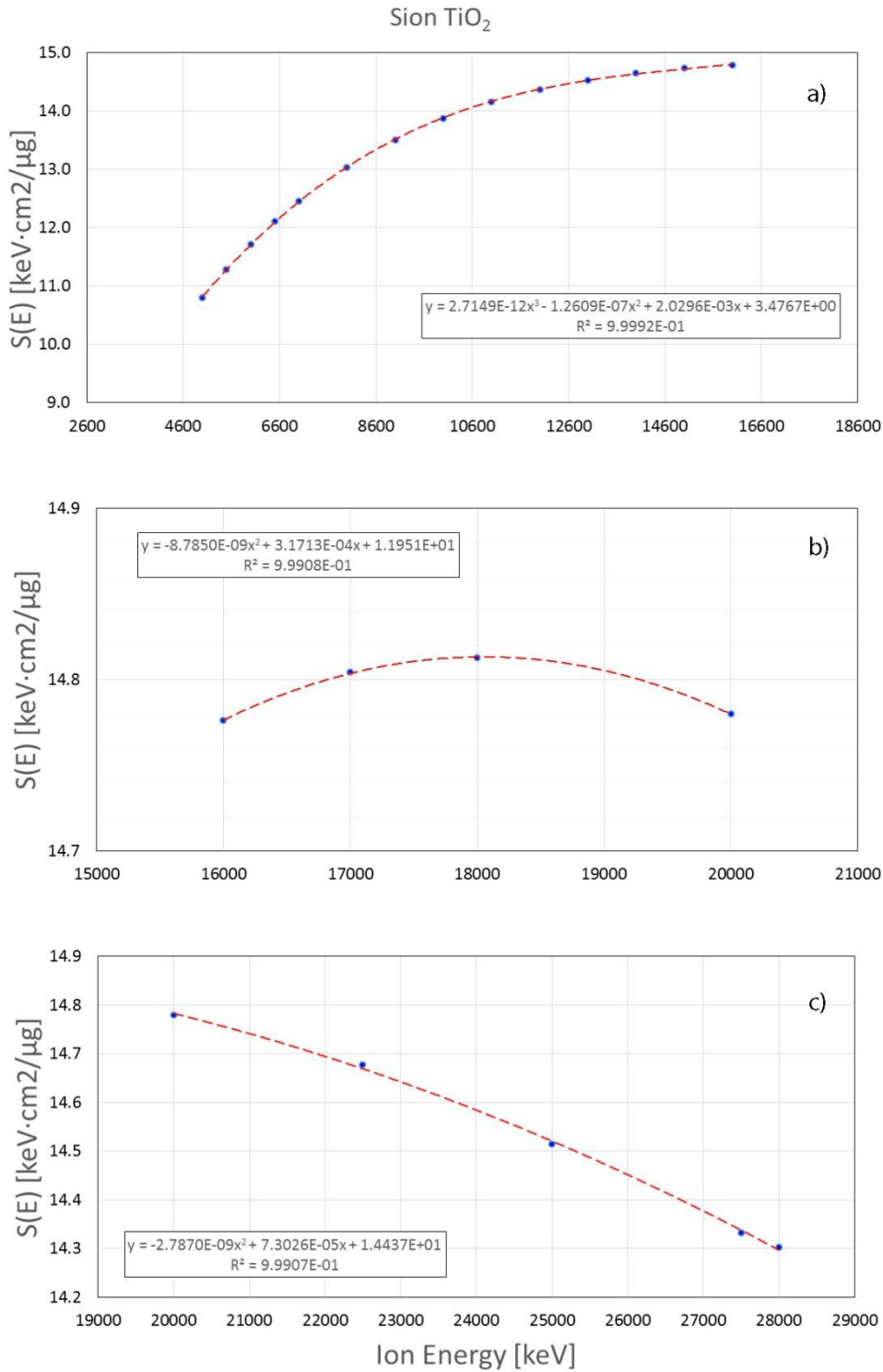

**Figure 1**
An example of the fitting of the specific energy loss curves *S(E)* in three energy ranges (a,b,c) for the case of Si ions interacting with a TiO$_2$ target. The three fits cover all the ranges starting at the beam energies listed in Table I and ending at the corresponding target exit energies.

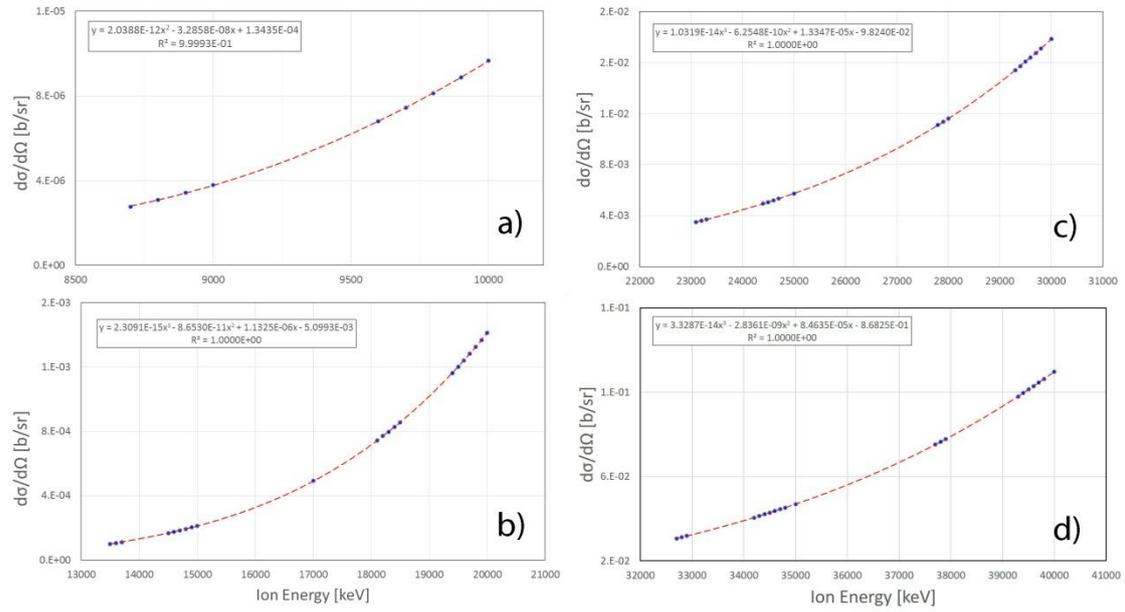

**Figure 2**
An example of the fitting of the ECPSSR cross sections $d\sigma/d\Omega(E)$ in four energy ranges (a,b,c,d) for the case of Si ions interaction with a $Fe_3O_4$ target. The four fits cover all ranges starting at the beam energies listed in Table I and ending at the corresponding target exit energies.

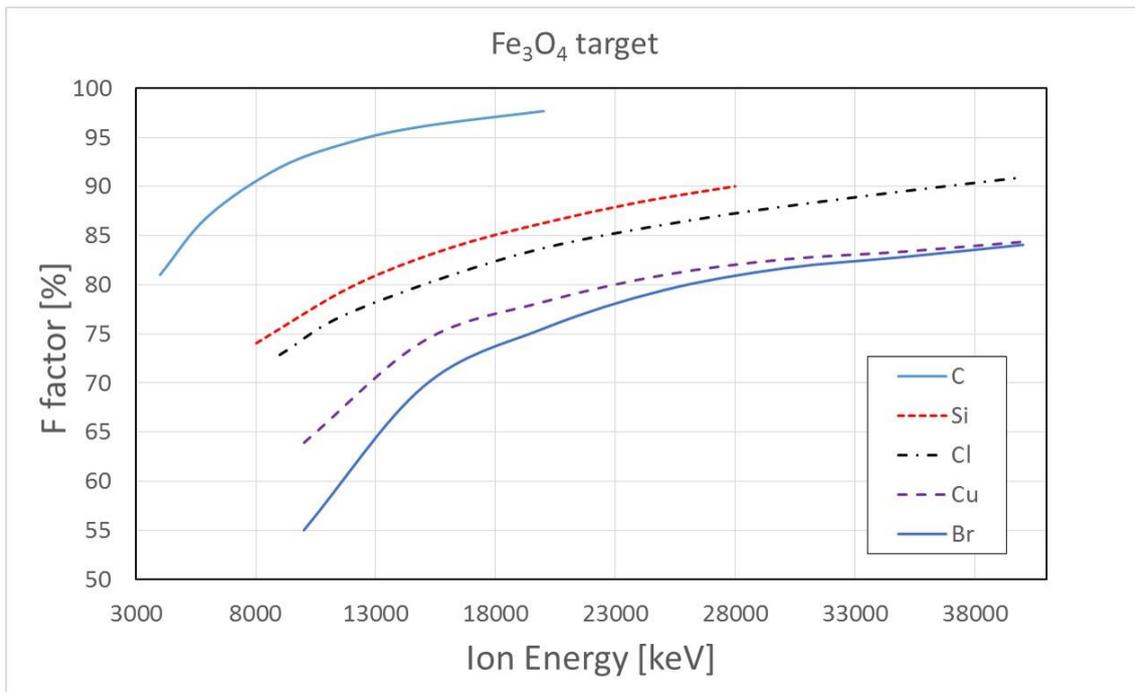

**Figure 3**
Correction factors *F* calculated for a $Fe_3O_4$ target bombarded by the following ions: C, Si, Cl, Cu and Br. For each ion the calculation has been performed up to the maximum beam energy listed in Table I.

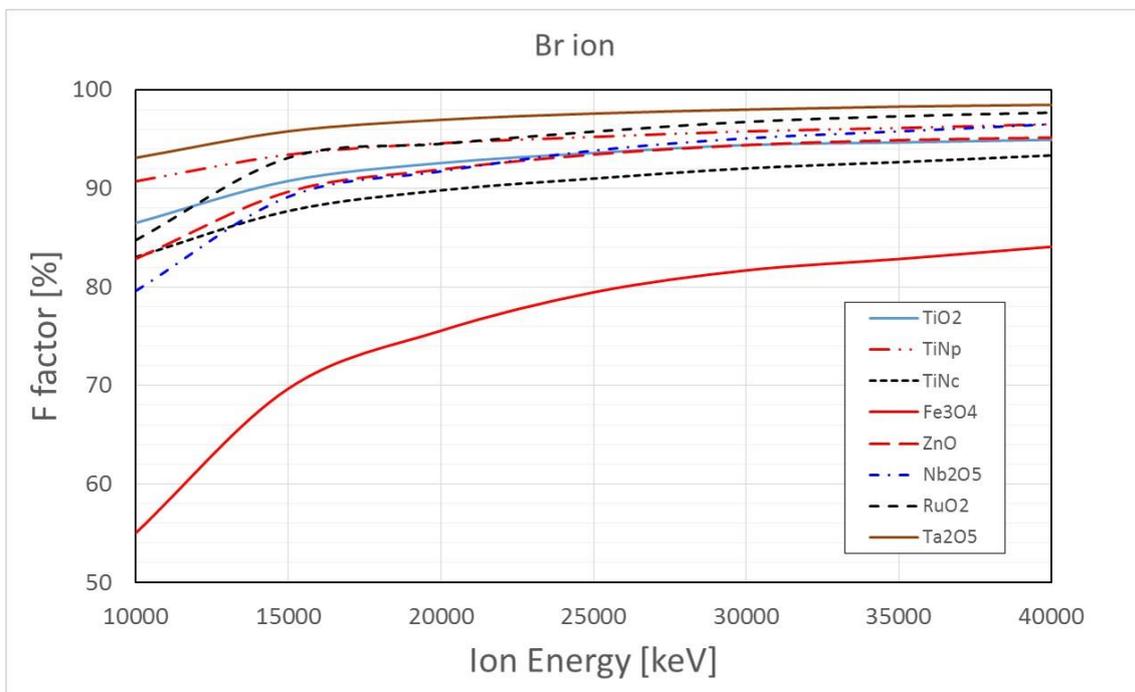

**Figure 4**  Correction factors *F* calculated for Br ions impinging on all the targets listed in Table I.